# Comment on "Comparing two formulations of skew distributions with special reference to model-based clustering" by A. Azzalini, R. Browne, M. Genton, and P. McNicholas

Geoffrey J. McLachlan, Sharon X. Lee
Department of Mathematics, University of Queensland,
St. Lucia, Brisbane, Australia

**Abstract**

In this paper, we comment on the recent comparison in Azzalini et al. (2014) of two different distributions proposed in the literature for the modelling of data that have asymmetric and possibly long-tailed clusters. They are referred to as the restricted and unrestricted skew $t$-distributions by Lee and McLachlan (2013a). Firstly, we wish to point out that in Lee and McLachlan (2014b), which preceded this comparison, it is shown how a distribution belonging to the broader class, the canonical fundamental skew $t$ (CFUST) class, can be fitted with essentially no additional computational effort than for the unrestricted distribution. The CFUST class includes the restricted and unrestricted distributions as special cases. Thus the user now has the option of letting the data decide as to which model is appropriate for their particular dataset. Secondly, we wish to identify several statements in the comparison by Azzalini et al. (2014) that demonstrate a serious misunderstanding of the reporting of results in Lee and McLachlan (2014a) on the relative performance of these two skew $t$-distributions. In particular, there is an apparent misunderstanding of the nomenclature that has been adopted to distinguish between these two models. Thirdly, we take the opportunity to report here that we have obtained improved fits, in some cases a marked improvement, for the unrestricted model for various cases corresponding to different combinations of the variables in the two real datasets that were used in Azzalini et al. (2014) to mount their claims on the relative superiority of the restricted and unrestricted models. For one case the misclassification rate of our fit under the unrestricted model is less than one third of their reported error rate. Our results thus reverse their claims on the ranking of the restricted and unrestricted models in such cases.

## 1 Introduction

In this paper, we provide some comments on the preprint, Azzalini et al. (2014), which we shall refer to as ABGM in the sequel. In ABGM a comparison is given of two different distributions proposed for the modelling of data that have asymmetric and possibly long-tailed clusters. They refer to the two models as the classical and SDB, the latter so named since it was proposed by Sahu, Dey, and Branco (2003). Previously, these two distributions were referred to as the restricted and unrestricted skew $t$-distributions by Lee and McLachlan (2013a). We shall continue to use this latter terminology in our comments below.

We firstly note that in the paper of Lee and McLachlan (2014b), which was posted on the arXiv.org e-Print archive before the appearance of ABGM there, it is shown how a distribution belonging to the broader class, the canonical fundamental skew $t$ (CFUST) class, can be fitted with essentially no additional computational effort than for the unrestricted distribution. The CFUST class includes the restricted and unrestricted distributions as special cases. We let FM-rMST, FM-uMST, and FM-CFUST refer to finite mixtures of multivariate restricted, unrestricted, and canonical fundamental skew $t$-distributions, respectively. With the availability of software for the fitting of the FM-CFUST model, users now have the option for letting the data decide as to which model is appropriate for their particular dataset. Or they can fit all three models (FM-rMST, FM-uMST, and FM-CFUST) and make their own choice between the three. The FM-CFUST model is to be defined in Section 3.

Our main purpose in writing this paper is to respond to statements in ABGM that are apparently based on a serious misunderstanding of the reporting of results in Lee and McLachlan (2014a) and, in particular, of the nomenclature used there. The discussion of our work in ABGM is limited to Lee and McLachlan (2014a), and so it does not consider the results presented in our other papers, in particular, Lee and McLachlan (2013a, 2013b, 2013c), although they are cited in ABGM. It is particularly unfortunate that these papers are not included in the comparison in ABGM as they contain a comparison of the restricted and unrestricted models applied to seven datasets from various fields. Also, explicit cautionary notes are made in them to guard against any potential misunderstanding of our terminology. For example, in Lee and McLachlan (2013a, Page 244) it is stated that "Note that the use of 'restricted' here refers to restrictions on the random vector in the (conditioning-type) stochastic definition of the skew distribution. It is not a restriction on the parameter space, and so a 'restricted' form of a skew distribution is not necessarily nested within its corresponding 'unrestricted' form." A more detailed response to the misrepresentation of our work is to be given in Section 6.

We take the opportunity to report here that we have obtained improved fits, in some cases a marked improvement, for the unrestricted model for the two real datasets that were used in ABGM to mount their claims on the relative superiority of the restricted and unrestricted models. For the particular two-variable combination that was the focus of their attention for their analysis of the crab dataset, the misclassification rate (MCR) of our fit under the unrestricted model is only 0.11 compared to their reported MCR of 0.36 (versus 0.15 for the restricted model).

For the other dataset considered in ABGM, the AIS data, our analysis found that the unrestricted model had a smaller MCR than the restricted for a slightly greater number of the bivariate and trivariate combinations of the 11 variables in the dataset. This is in contrast to the finding in ABGM (Page 13) that "there are many more pairs and triplets of the 11 available variables for which the classical formulation outperforms the SDB formulation (cf. Figure 9)."

Before we give the improved fits for the unrestricted skew $t$-mixture model and respond to some of the statements in ABGM in which our work, in particular our nomenclature, is misrepresented, we shall give briefly a short history on how we came to be interested in fitting mixtures of skew $t$-distributions. Most importantly, we wish to stress that the reporting of our comparisons of the restricted and unrestricted models in our published work has been concerned solely with their relative behaviour on the datasets that we have analysed. No extrapolation to a general scenario is implied since the restricted model is not nested within the unrestricted model. But that is not to say that one of the models might have wider applicability than the other, and this is why we have reported our analyses for a wide variety of datasets.

# 2 Our introduction to mixtures of skew $t$-distributions via flow cytometry

In an attempt to provide an automated approach to the clustering of flow cytometry data, Pyne et al. (2009) considered the fitting of skew $t$-distributions. Skew component distributions were considered because the clusters tended to be asymmetric (non-elliptical in shape) and the $t$- rather than the normal versions of the skew distributions were adopted as the clusters tended to be long-tailed. Initially, Pyne et al. (2009) considered mixtures of skew $t$-distributions that belonged to the family of skew $t$-distributions proposed by Sahu et al. (2003).

Members of the latter family have the following characterization. The $p \times 1$ random vector $\boldsymbol{Y}$ can be expressed as

$$\boldsymbol{Y} = \boldsymbol{\mu} + \boldsymbol{\Delta}|\boldsymbol{U}_0| + \boldsymbol{U}_1, \tag{1}$$

where

$$\begin{bmatrix} \boldsymbol{U}_0 \\ \boldsymbol{U}_1 \end{bmatrix} \sim N_{2p}\left(\begin{bmatrix} \boldsymbol{0} \\ \boldsymbol{0} \end{bmatrix}, \frac{1}{w}\begin{bmatrix} \boldsymbol{I}_p & \boldsymbol{0} \\ \boldsymbol{0} & \boldsymbol{\Sigma} \end{bmatrix}\right). \tag{2}$$

In the above, $\boldsymbol{\mu}$ is a $p$-dimensional vector, $\boldsymbol{\Delta}$ is a $p \times p$ diagonal matrix, $\boldsymbol{I}_p$ denotes the $p \times p$ identity matrix, $\boldsymbol{\Sigma}$ is a positive definite matrix, and $\boldsymbol{0}$ is a vector/matrix of zeros with appropriate dimensions. Also, $w$ is the realization of the random variable $W$ distributed as gamma$(\frac{\nu}{2}, \frac{\nu}{2})$, and $|\boldsymbol{U}_0|$ denotes the vector whose $i$th element is the magnitude of the $i$th element of the vector $\boldsymbol{U}_0$.

In order to simplify the application of the EM algorithm to fit mixtures of these skew $t$-distributions, Pyne et al. (2009) imposed the restriction

$$U_{01} = U_{02} = \ldots = U_{0p} \tag{3}$$

on the $p$ latent skewing variables, where $U_{0i} = (\boldsymbol{U}_0)_i$ ($i = 1, \ldots, p$). This produces a distribution equivalent to the skew $t$-distribution formulated by Branco and Dey (2001) and Azzalini and Capitanio (2003) after reparameterization. Lee and McLachlan (2013a, 2014a) termed this distribution the restricted skew $t$-distribution to distinguish it from the distribution proposed by Sahu et al. (2003). For ease of reference, they termed the latter the unrestricted skew $t$-distribution since it can be characterized without any restrictions on the $p$ latent skewing variables in the convolution-type stochastic representation (1).

Although we found that finite mixtures of restricted multivariate skew $t$-distributions (FM-rMST) provided a good fit to our flow cytometry datasets, two- and three-dimensional plots of the markers on the cells suggested that the fit could be improved. With Lee and McLachlan (2014a) showing how the E-step of the EM algorithm can be implemented in closed form (apart from the updates of the component degrees of freedom), we started to fit finite mixtures of unrestricted multivariate skew $t$-distributions (FM-uMST) to datasets in flow cytometry. In our initial applications, we found the unrestricted model to give a lower MCR than the restricted model FM-rMST for the clustering of flow cytometry datasets where the cluster labels were compared to labels from experts obtained via manual gating. For example, the MCR of the restricted model was reduced by approximately 58%, 23%, and 136% in three separate flow cytometry datasets analysed in Lee and McLachlan (2013a), Lee and McLachlan (2013b), and Lee and McLachlan (2014a), respectively. Since then we have continued to analyse datasets taken from the flowCAP1 competition in flow cytometry (Aghaeepour et al., 2013), where the labels from experts are available.

# 3  Canonical fundamental skew $t$-distribution

We would like to point out that in our preprint Lee and McLachlan (2014b) that preceded ABGM, we have provided the EM equations for the fitting of a mixture of CFUST (canonical fundamental skew $t$) distributions. The CFUST distribution was introduced (and so named) as a canonical version (special case) of the fundamental skew $t$-distribution by Arellano-Valle and Genton (2005). The attractive feature of the CFUST model is that it includes what we call the restricted and unrestricted distributions as special cases. And it can be fitted with essentially no extra effort over the fitting of the unrestricted normal or skew $t$-distributions. We had not realized that generalizing the $p \times p$ diagonal matrix $\mathbf{\Delta}$ of skewness parameters to a non-diagonal $p \times q$ matrix in the convolution definition would require no extra effort in calculating the EM equations until we attempted such computations late last year.

The unrestricted skew $t$-distribution can be obtained from the CFUST family by setting $q = p$ in the skewness matrix $\mathbf{\Delta}$ and taking it to be diagonal. The restricted version can be obtained by either setting $q = 1$ or by setting $q = p$ and taking all the elements of $\mathbf{\Delta}$ to be zero except for those in one column.

In our fitting of finite mixtures of (multivariate) CFUST distributions, we have tended to use the same starting strategy as for the unrestricted model. That is, we have taken $\mathbf{\Delta}$ to be diagonal on the first iteration for each component.

# 4  Unrestricted versus restricted skew $t$-distribution

As the restricted skew $t$-distribution is not nested within the unrestricted skew $t$-family, one can generate datasets where one will be preferable to the other. Given this, we do not see how one can try to establish that the restricted and unrestricted models will give, say, comparable results in general by consideration of only a few real datasets. For example, only two real datasets are considered in ABGM.

In our approach to working with these models, we usually fit mixtures of restricted skew $t$-distributions in the first instance given that they can be fitted much more quickly than the unrestricted version. In proceeding then to fit mixtures of unrestricted skew $t$-component distributions, we use the clustering provided by the restricted fit as one of our initial partitions of the data. This is in addition to using random, $k$-means, and (ordinary) $t$-mixtures based starts. And now we have available software (Lee and McLachlan, 2014b) for the fitting of FM-CFUST, a wider class that includes the restricted and unrestricted distributions as special cases. Thus our approach is to provide the methodology and software for users to have the options to fit the various models and to make their own choice on the basis of the results that are so obtained for their particular dataset.

It is remarked in ABGM (Page 4) that "LM2012 [that is, Lee and McLachlan (2014a)] only consider examples with $d \leq 4$," where $d$ refers to the number of variables $p$ in our notation here. This is not so. In Lee and McLachlan (2014a, Section 7), we actually fitted the FM-uMST model in an example with $p=10$ variables, and in the analysis of the AIS dataset discussed in this paper, we fitted this model to all $p=11$ variables. Of course, as in any multivariate setting with arbitrary covariance (scale) matrices, one needs to pay attention to the relative size of $p$ to the sample size $n$, as $p$ grows in size.

# 5 Existence of improved fits by unrestricted skew $t$-mixtures

In ABGM, three datasets are considered. The first is a simulation one which we shall not comment on as we do not know the parameters of the distribution from which the data were generated and the class labels. Thus in the sequel, our remarks are confined to the two real datasets.

Concerning the crab dataset analysed in ABGM, the correlations between any two of the five variables is very high (the lowest is 0.89). So this limits the unrestricted model producing a much better fit than the restricted model. More precisely, in the context of bivariate combinations of the variables for these data, it effectively means that bivariate datasets will lie almost on a straight line and so the restricted model with its univariate skewing function should not be disadvantaged. Thus on bivariate datasets of the crab dataset, the restricted model cannot perform too far below that of the unrestricted model and may well be much better for some.

For the crab dataset, the focus in ABGM is on the two-variable case using the variables RW and FL, for which it is found that the restricted model is preferable to the unrestricted. It is stated in ABGM (Page 10) that "...the classical skew-$t$ mixture model gives good classification performance with only 30 misclassifications, corresponding to an ARI of 0.487. The SDB skew-$t$ mixture model (that is, the unrestricted model), however, gives very poor classification, producing results only slightly better than would be expected under random classification (ARI=0.074), with 72 of the 200 crabs misclassified."

We had not previously fitted skew mixture models to this dataset. On now analysing this dataset, we have found that for the bivariate case with the RW and FL variables, the unrestricted model gives a clustering with only 22, not 72, misclassifications; that is, the MCR is only 0.11 (22/200) compared to 0.15 (30/200) for the restricted model. Our analysis did find a clustering with a slightly smaller MCR of 0.125 (25/200) for the restricted model.

In ABGM (Page 13), it is stated that "if one really wants to rank them, the classical version (that is, the restricted) performed slightly better for the crab data." However, on our subsequent analysis of all datasets, we have found that the unrestricted version performs slightly better than the restricted. For example, on considering all 26 datasets corresponding to the 26 different combinations of the five variables, we found that the restricted model gave a better fit for only 3 of the 26 datasets. The differences were generally small with there being 13 ties. We modified the EMMIXskew and EMMIXuskew packages so that the two models could be started using the same values.

As for the AIS dataset, it is not surprising that the performances of the restricted and unrestricted models are quite similar for the two- and three-dimensional combinations of the variables considered in ABGM, particularly for the bivariate combinations of the variables that have high correlations. On considering all 220 datasets corresponding to the 220 combinations consisting of all pairs and triplets of the 11 variables, we found that the unrestricted model gave a better fit for 105 versus 100 combinations for the restricted. The differences were generally small with there being 15 ties. But it is in contrast to the result reported in ABGM (Page 13), which states that "there are many more pairs and triples for which the classical formulation outperforms the SDB formulation (cf. Figure 9)." It is worth noting that if we use all 11 variables available, then the unrestricted model gives a MCR of 0.0198 compared to 0.0297 for the restricted model (that is, two fewer misallocations).

In ABGM (Page 8), it is stated that "it is the purpose of the remaining part of this section to explore in a more systematic and fairer way the performance of the two forms of skew-$t$ distributions in clustering applications. An important aspect in these comparisons is the potential bias due to selective reporting." Our reason for analysing the bivariate data on the Ht and Bfat variables in the AIS dataset in Lee and McLachlan (2014a, 2013c) was to demonstrate the potential gain that can be achieved even in low dimensions by using the unrestricted model, which had a MCR of 0.0941 (19/202) versus 0.2228 (45/202) for the restricted model (Lee and McLachlan, 2013c, Page 14). There was no suggestion that this relative superiority of the unrestricted model would necessarily be maintained for all combinations of the 11 variables in the AIS data, and so the questions of fairness of comparison and selection bias do not arise.

# 6 Response to some statements in ABGM

In ABGM, a number of statements are made that demonstrate that there is a serious misunderstanding of the work reported in Lee and McLachlan (2014a), in particular, of the nomenclature that they used. In this section, we respond briefly to some of these statements. Also, we note that the misunderstanding is not limited to the statements explicitly considered below.

Before we proceed to list explicitly some of these statements, it is clear that ABGM is caught up in nomenclature. It is stated in ABGM (Page 13) that "An argument made throughout this paper is that the nomenclature 'restricted' and 'unrestricted' is inappropriate." Ideally, it would have been better if we had managed to come up with a nomenclature that was universally acceptable, assuming that such a goal were achievable. However, our nomenclature is not inappropriate.

Against this background, we now proceed to address some of the statements in ABGM where we consider that the reporting of the results in Lee and McLachlan (2014a) is misrepresented.

(1) **Statement 1 (Page 7 of ABGM):** *In these expositions the skew-normal and skew-t distributions which we refer to as classical are named 'restricted' and those that we refer to as SDB are named 'unrestricted,' under the incorrect assumption that the SDB variants constitute a more general family than the classical ones.*

   As explained in the introduction, we called the model of Sahu et al. (2003) the unrestricted model since it did not place any restrictions on the $p$ latent skewing variables $U_1, \ldots, U_p$.

   In Lee and McLachlan (2013b), we say in the last two lines of Page 431 that "It should be stressed that the rMSE family and uMSE family match only in the univariate case, and one cannot obtain (7) [the restricted distribution] from (9) [the unrestricted distribution] when $p > 1$.

   Furthermore, in Lee and McLachlan (2013a), we say on Page 244 that "Note that the use of 'restricted' here refers to restrictions on the random vector in the (conditioning-type) stochastic definition of the skew distribution. It is not a restriction on the parameter space, and so a 'restricted' form of a skew distribution is not necessarily nested within its corresponding "unrestricted" form. Then in our concluding remarks in the same paper on Page 264, we say, "Concerning the use of the terminology 'restricted' and 'unrestricted', it should be noted that the restricted skew forms are not nested within the corresponding unrestricted forms, ..."

(2) **Statement 2 (Page 7 of ABGM):** *Not only does the adopted terminology, restricted vs. unrestricted, convey a message of broader generality in the second case, but this is also explicit in the use of the term 'extension,' which is inappropriate because we have seen earlier that neither one of the families is a subset of the other for $(d > 1)$.*

We used the word "extension" to describe how the unrestricted skew $t$-distribution can be obtained by an "extension" of the convolution-type formulation of the restricted version in which the single (latent) skewing variable is replaced by a multivariate random variable (and similarly with the conditioning-type formulation). This use of "extension" does not necessarily imply that the restricted distribution is a special case of the unrestricted version.

(3) **Statement 3 (Page 8 of ABGM):** *LM2012 [Lee and McLachlan (2014))] contains claims like "the superiority of FM-UMST model is evident (end of Section 6.1) and ... (end of Section 6.2). These comments refer to two specific cases, but are taken as the basis for the general statement 'Examples on several real datasets shows shows [sic] that the unrestricted model is capable of achieving better clustering than the restricted model (Section 8)."*

In our reporting of results of fitting the restricted and unrestricted models, we have qualified our statements by saying something like "in the present dataset." For instance, in Lee and McLachlan (2014a) we say in the last lines of Section 6.1 on Page 194 that "Thus the superiority of the FM-uMST model is evident in dealing with the asymmetric and heavily tailed data **in this dataset**." Then near the end of Section 6.2, we say "The FM-rMST model has a disappointing performance in terms of clustering **for this dataset** ..." Similarly, in Lee and McLachlan (2013a) on Page 264, we write "The results from Table 12 reveal that the unrestricted model is more accurate than the restricted variant **for this dataset.**"

Concerning our statement in Section of 8 of Lee and McLachlan (2014a) that "Examples on several real datasets show that the unrestricted model is capable of achieving better clustering than the restricted model," we were conveying our experience at the time based on (a) our fitting of these two models to the two real datasets in Lee and McLachlan (2014a), where we found that the unrestricted model FM-uMST model provided a noticeably better fit than the restricted model in the Lymphoma data and gave a lower MCR than the restricted model in the GvHD data; (b) our unpublished analyses of several real datasets, in which it was found that the unrestricted model gave improved clustering results in comparison to the restricted model. The latter were subsequently published in Lee and McLachlan (2013a,, 2013b, 2013c, 2013d), and McLachlan and Leemaqz (2013)).

(4) **Statement 4 (Page 7 of ABGM):** *The authors state that the "form of skewness is limited in these characterizations. In Sect. 5 we study an extension of their approach to the more general form of skew t-density as proposed by Sahu et al. (2003)." This claim of limited form of skewness is completely unsupported: no expressions of any measure of skewness is even reported.*

As explained in Lee and McLachlan (2014a), the restricted model has a univariate skewing function which is a limitation. We are not saying there is a limit on the amount of skewness under the restricted model; rather the limitation refers to having only a univariate skewing function. This limits its capacity to handle certain multivariate skew data. A specific example of the latter is the case of independent skew variables.

# 7 Concluding remarks

We have provided some comments on the comparison in ABGM of what have been termed in Lee and McLachlan (2013a) as the restricted and unrestricted skew $t$-distributions. As mentioned in Section 2, our interest in the unrestricted skew $t$-mixture model was motivated by our search to find a model that gave an improved fit over the restricted model for datasets from flow cytometry. We have subsequently investigated its applicability in datasets from a wide variety of fields.

In this work, our interest has been on the development of the methodology for the fitting of skew symmetric mixtures with components that have multivariate skewing functions. Given the availability of this option to a user, we have been interested to assess the potential of this more computationally demanding model to provide an improved fit over existing models. It is realized that the unrestricted model need not provide an improved fit over the restricted model in any one given dataset.

We now have developed an algorithm that fits a mixture of CFUST distributions that include both restricted and unrestricted distributions as special cases (Lee and McLachlan, 2014b). Thus the user has the option of fitting the restricted and unrestricted distributions or proceeding directly with a more flexible distribution that includes the former two as special cases.

We have been keen to explain in our comments on the comparison in ABGM how several statements there misrepresent our reporting of results in the area. In particular, it is evident there has been a misunderstanding of the nomenclature that we have used to differentiate between the models in our references to them. We are somewhat surprised that this misunderstanding of our nomenclature still exists given that we have explicitly explained its use in our papers such as Lee and McLachlan (2013a, 2013b).

We have also taken the opportunity to report that we have found a much improved clustering performance of the unrestricted mixture model compared to that reported in ABGM for the crab dataset. For the particular two-variable combination that was the focus of their analysis of this dataset, we found a clustering produced by the unrestricted model with a MCR that was less than one third of that reported in ABGM for this model. Concerning the other dataset in ABGM that was used to support their position, the AIS dataset, we have found for the two- and three-dimensional combinations of the variables in this dataset considered in ABGM that the unrestricted and restricted models perform very similarly but with the former slightly shading the restricted. This is in contrast to the result reported in ABGM where it is stated that there are many more pairs and triples for which the restricted model outperformed the unrestricted.